\documentclass[prd,aps,amsmath,amssymb,letter,showpacs]{revtex4}
\usepackage{graphicx}% Include figure files
\usepackage{epsfig,rotate,latexsym}
\usepackage{hyperref}
\usepackage{subfigure}
\begin{document}
\title{Z(3) metastable states in Polyakov Quark Meson model}
\author{Ranjita K. Mohapatra}
\email {ranjita.iop@gmail.com}
\author{Hiranmaya Mishra}
\email {hm@prl.res.in}

\affiliation{Theory Division, Physical Research Laboratory, Navrangpura, 
Ahmedabad 380009, India}

\begin{abstract}

We study the existence of Z(3) metastable states in the presence of the 
dynamical quarks within the ambit of Polyakov quark meson (PQM) model. 
Within the parameters of the model, it is seen that for temperatures T$_m$ 
greater than the chiral transition temperature T$_c$, Z(3) metastable 
states exist ( $\text T_{m} \sim 310$ MeV at zero chemical potential). At finite chemical 
potential $\text T_m$  is larger than the same at vanishing chemical potential. We also observe a 
shift of ($\sim 5^\circ$) in the phase of the metastable vacua at 
zero chemical potential. The energy density difference between true and Z(3) metastable 
vacua is very large in this model. This indicates a strong explicit symmetry breaking 
effect due to quarks in PQM model. We compare this explicit symmetry breaking in PQM
model with small explicit symmetry breaking as a linear term in Polyakov loop 
added to the Polyakov loop potential. We also study about the possibility of domain 
growth in a quenched transition to QGP in relativistic heavy ion collisions.    

\end{abstract}


\maketitle

\section{INTRODUCTION}

The structure of QCD vacuum and its modification under extreme environment has been 
a major theoretical and experimental challenge in strong interaction physics. 
Heavy ion collisions at Relativistic Heavy Ion Collider (RHIC) and at Large Hadron 
Collider ( LHC ) provide an opportunity to investigate the modification of the vacuum
structure of QCD as related to nonperturbative aspects of QCD. There
has been  strong evidence that a strongly interacting Quark Gluon Plasma 
(QGP) is produced in these experiments. With the increase in collision 
center of mass energy from RHIC to LHC, a QGP state at higher temperature
$\sim 2\text T_c$ is expected to be formed at LHC \cite{wilde}. At such  
high temperature, it is important to study the non-trivial vacuum 
structure of QCD arising from Z(3) center symmetry of QCD \cite{plkv}. 
In pure $SU(3)$ gauge theory, that can be considered as QCD with infinitely heavy quark masses, 
the confining phase is center symmetric with vanishing expectation value of the Polyakov 
loop order parameter. On the other hand, the deconfinement is characterized by a non vanishing
value of the order parameter with three degenerate vacua corresponding to three 
different phases of Z(3) center symmetry. However, for real QCD
with inclusion of dynamical quarks, Z(3) center symmetry is 
explicitly broken in the deconfined phase  with one true
vacuum  and two metastable vacua \cite{dixit,weiss,ogilvie96}. It is generally believed that the 
explicit symmetry breaking effect due to quarks is  small 
and a linear term in Polyakov loop is added to Polyakov loop potential
to take into account of this effect \cite{psrsk2,latic1}. For these models,
metastable states exist at any temperature greater than $\text T_c$. However, some recent lattice 
QCD result \cite{deka} show that these metastable states do not exist in 
the neighborhood of $\text T_c$($\sim 200$ MeV), but for temperatures $\text T\ge 750$ MeV. This leads to a 
strong explicit symmetry breaking rather than small explicit symmetry breaking due to quarks.

It has been studied that these Z(3) domains give a 
microscopic explanation for large color opacity ( jet quenching) and 
near perfect fluidity ( small value of $\eta/s$) nature of QGP 
\cite{bass,monnai}. Hence, it is important to study these domains
in the nonperturbative regime of QCD in which the system exists in a 
strongly interacting QGP phase just 
after the collision of two heavy nuclei in relativistic heavy ion collision. 
Due to the explicit symmetry effect of quarks, there are huge 
oscillations in the Polyakov loop order parameter field which gives rise
to large fluctuations in the flow anisotropies in the quenched transition
to QGP \cite{ranj}.  The importance of Z(3) walls has also been discussed as
non-trivial scattering of quarks from Z(3) walls. Its consequences 
for cosmology as well as for heavy ion collision have been explored 
\cite{layek,gupta,gupta2,ananta}, including
the possibility of CP violating scattering of quarks from Z(3) walls
leading to interesting observational implications \cite{atreya}.

These Z(3) metastable states have been studied at high temperatures
in presence of dynamical quarks \cite{dixit,ogilvie96}. It has been 
shown that the contribution of massless quarks to the one loop 
effective potential leads to metastable states and the free energy density difference
between true and metastable vacua is given by $\frac{2}{3}
\pi^2\text T^4\frac{\text N_{f}}{\text N^3}(\text N^2-2)$, $\text N_f$, $\text N$ being the number of light flavors 
and number of colors respectively.
 Since the calculation of the 
effective potential here is perturbative, it is only valid at temperatures 
much larger than $\Lambda_{QCD}$. We do not expect it to be valid 
near $\text T_c$. However, a detailed description of such metastable states 
near $\text T_c$ is difficult both because of the non perturbative nature of QCD and 
the incompleteness of the theory of thermodynamics of the non equilibrium systems.

Therefore, in the present work, we examine these metastable states 
near $\text T_c$ using Polyakov Quark Meson (PQM) model \cite{scfer}. 
This model is based on the two important aspects of QCD phase transition, 
chiral transition and confinement-deconfinement phase transition. 
The Polyakov loop potential represents 
SU(3) pure gauge theory part which respects Z(3) center symmetry.
The linear sigma model is included to represent the chiral
symmetry of QCD. The quarks are minimally coupled to a spatially constant
temporal background gauge field $A_0$. The model thus cleverly uses the chiral 
as well as confining properties of QCD. Since the constituent quark masses are much 
larger than the mass of pions, the meson dynamics dominate at low temperatures 
reproducing the results of chiral perturbation theory. Since, this model describes an 
interaction potential between quarks, mesons and Polyakov loop, this is 
a very suitable model to study the metastable states near $\text T_c$.     

It has been shown that the sign of finite temperature correction to the
effective potential depends on the sign of real part of the Polyakov loop
\cite{ogilvie96}. In general, it gives a negative contribution to the energy
density of the vacuum when the real part of Polyakov loop is positive 
( making it a global minimum for $\theta=0$). However, 
it gives a positive contribution to the effective potential (making it 
metastable) when the real part of the Polyakov loop is negative. This is also the 
case for PQM model. The Polyakov loop potential respects Z(3) symmetry and 
there are three degenerate vacua at any temperature greater than the 
critical temperature. However, when the interaction term 
between Polyakov loop and quarks is included as given in PQM model, $\theta=0$ vacuum 
becomes true vacuum due to addition of a negative energy density since the real part
of Polyakov loop is positive. The other two Z(3) vacua become metastable due to addition 
of a positive energy density since the real part of Polyakov loop is negative.
Since the interaction part between the Polyakov 
loop and quarks in PQM model also has chemical potential dependence, we study
about these metastable states and their temperature and chemical potential
dependence in great detail in this model.  In this paper, we also have shown
that, of different forms of the Polyakov loop potentials those with large barrier 
between different Z(3) vacuum, sustain Z(3) metastability structure when the 
interaction term between quarks and Polyakov loop is added.      

It is important and relevant to study non equilibrium effects since 
there are metastable states in the system.
In classical non-equilibrium thermodynamics, metastable states play a crucial role in
explaining various phenomena like
super cooling and super saturation. In analogy, one might expect interesting 
phenomenology of topological 
super cooling of QGP in heavy ion collision. This has been already 
studied as a numerical simulation in Ref. \cite{ranj} where the phase 
transition from confined to deconfined phase is modeled by a quench to 
a very high temperature with in Polyakov loop potential. The evolution 
of Z(3) domains have been studied for different cases like small or
large explicit symmetry breaking effect by putting 
a small or large term as the coefficient of linear term in the 
Polyakov loop potential\cite{ranj}. In  present investigation, we discuss 
domain growth for the Polyakov loop and  quark condensate  order parameter in a 
quench scenario. The non equilibrium effects have been studied using 
Langevin equations  within different effective models of QCD like 
PQM model \cite{herold} as well as Nambu-Jona-Lasinio (NJL)
model \cite{purihm}.

 The paper is organized in the following manner. In section II,
we discuss the essential aspects of the PQM model. 
Here, we discuss about the two different parameterization used in the
literature for Polyakov loop potential in SU(3) pure gauge theory.
Section III describes about the metastable states at higher temperature and 
chemical potential. It is shown that the parameterization for the pure gauge part 
which has larger barrier between different vacua leads to existence of  
metastable states when quarks are coupled to it.
We show that there is a large symmetry breaking effect due to quarks
which leads to the existence of metastable states at higher temperature 
and a large shift ($\sim 5^\circ$) in the phase of 
metastable vacua. We compare the explicit symmetry breaking of PQM model with
small explicit symmetry breaking as a linear term in Polyakov loop added to Polyakov
loop potential. Sec. IV discusses the numerical techniques for the
simulation to describe phase transition via quench and the results of the simulation.
 Finally, in section V, we summaries the results of the present investigation 
and give a possible outlook.

\section{THE POLYAKOV-QUARK-MESON MODEL}

\subsection{ The Polyakov loop potential}

The thermal expectation value of the Polyakov loop represents the 
order parameter for the confinement-deconfinement phase transition. 
The Polyakov loop operator is a Wilson loop in the temporal direction

\begin{equation}
{\cal{P}} = P \exp\left( i \int_0^\beta {dx_0 A_0(x_0)}\right),
\label{eq0}
\end{equation}
where, $A_0(x_0)$ is the temporal component of the gauge field 
$A_{\mu}$, $P$ denotes path ordering in the Euclidean time $\tau$, 
and $\beta = 1/\text T$ with T being the temperature. Within Polyakov gauge,
the temporal component of the gauge field is time independent, 
 so that
 ${\cal{P}}=\exp ({ i \beta A_0})$.
Further, one can rotate the gauge field in the Cartan sub algebra 
$A_0^c=A_0^3 \lambda_3+A_0^8 \lambda_8$ .
The normalized Polyakov loop variable $\Phi$ and its charge conjugate
$\bar{\Phi}$ are defined as the 
color trace of the Polyakov loop operator defined in Eq.(\ref{eq0}).

\begin{eqnarray}
 \Phi = \frac{1}{N_c} tr {\cal{P}},\quad \bar{\Phi}=
\frac{1}{N_c} tr {\cal{P}}^\dagger
\label{eq1}
\end{eqnarray} 

Here the trace is taken in the fundamental representation. $\Phi$ and
$\bar\Phi$ are complex scalar fields and their mean values are related 
to the free energy of infinite heavy, static quark or anti-quark. The
order parameter $\Phi$ vanishes in the confined phase since infinite 
amount of free energy is required to put a static quark in that phase.
However, the order parameter is finite in deconfined phase related 
to finite free energy of static quark.

Under Z(N) center symmetry of SU(N) gauge symmetry, the Polyakov
loop order parameter transforms as 

\begin{eqnarray}
\Phi\to z\Phi,\quad z \in Z(N)\  
\end{eqnarray}

In pure SU(3) gauge theory which is the limit of QCD with infinitely heavy quarks, 
the confining phase is center symmetric so that $\langle\Phi\rangle=0$, whereas 
deconfinement is characterized by a non vanishing value of the Polyakov
loop expectation value, since center symmetry is broken spontaneously. In 
physical world, with finite quark masses  this symmetry is explicitly broken.

For SU(3) pure gauge theory, the effective potential of Polyakov loop
has been proposed with the parameters fitted to reproduce lattice results
for pressure and energy density.  Let us note here that the explicit form of the
Polyakov loop  is not known directly from first principle calculations. The approach 
has been to choose a functional form that reproduces crucial features of pure gauge 
theory adjust the parameters of the function so as to reproduce the thermo dynamical
observables of lattice simulations. Here, we will discuss
two kinds of parameterization used in literature as discussed below.
The parameter set 1 is taken from Ref. \cite{psrsk2} and the parameter 
set 2 is taken from Ref. \cite{rati,scfer}. Both the potential with the parameters 
given below represents a first order phase transition at the critical 
temperature $\text T_0 = 270$ MeV. For the parameter set 1, the effective Polyakov 
potential is given by

\begin{eqnarray}
\label{set1v}
&&\hspace{-1.4cm}\frac{{\cal U}(\Phi,\bar\Phi)}{\text T^4}=\left(-\frac{b_2}{4}
\left(|\Phi|^2+|\bar\Phi|^2 \right)
-\frac{b_3}{6}(\Phi^3+\bar\Phi^3)+\frac{1}{16}
\left(|\Phi|^2+|\bar\Phi|^2\right)^2\right)*b_4
\label{efpot1}
\end{eqnarray}

\noindent The modulus of $\Phi$ and $\bar\Phi$ are same for pure gauge theory. By writing 
$\Phi=|\Phi|e^{i\theta}$, one can see that the $b_3$ term  in Eq.(\ref{efpot1}) 
gives $\cos 3\theta$ term leading to Z(3) degenerate vacua for non vanishing value for $|\Phi|$
i.e. for temperatures greater than the critical temperature $\text T_0$.
The coefficients $b_3$ and $b_4$ have been taken as, $b_3=2.0$ and
$b_4=0.6016$. The temperature dependent coefficient $b_2$ 

\begin{equation}
b_2(\text T) = (1-1.11*\text T_0/\text T)(1+0.265*\text T_0/\text T)^2(1+0.3*\text T_0/\text T)^3-0.487
\end{equation}

\noindent With the coefficients chosen as above, the expectation
value of order parameter approaches to
$ x =b_3/2+\frac{1}{2} \sqrt{b_3^2+4b_2(\text T=\infty)}$ for temperature
$\text T \rightarrow \infty$. We use
the normalization such that the expectation value of order 
parameter $\Phi$ goes to unity for temperature $\text T \rightarrow \infty$. 
Hence the fields and the coefficients in the above potential are 
rescaled as $ \Phi \rightarrow \Phi /x$,
$b_2(\text T)\rightarrow b_2(\text T)/x^2$, $b_3\rightarrow b_3/x$ and
$b_4\rightarrow b_4x^4$ to get proper normalization of $\Phi$.

\noindent For the parameter set 2, the effective Polyakov potential is

\begin{eqnarray}
\label{set2v}
&&\hspace{-1.4cm}\frac{{\cal U}(\Phi,\bar\Phi)}{\text T^4}=-\frac{b_2}{4}
\left(|\Phi|^2+|\bar\Phi|^2 \right)
-\frac{b_3}{6}(\Phi^3+\bar\Phi^3)+\frac{b_4}{16}
\left(|\Phi|^2+|\bar\Phi|^2\right)^2
\end{eqnarray}

\noindent Here, the temperature independent coefficients $b_3 = 0.75$
, $b_4 = 7.5$ and temperature dependent coefficient $b_2$ is given by
 
\begin{equation}
 b_2(\text T) = a_0  + a_1 \left(\frac{\text T_0}{\text T}\right) + a_2
  \left(\frac{\text T_0}{\text T}\right)^2 + a_3 \left(\frac{\text T_0}{\text T}\right)^3
\end{equation}

\noindent where,  $a_0 = 6.75$, $a_1 = -1.95$, $a_2 = 2.625$, $a_3 = -7.44$.

The difference between these two parameterization is that the 
barrier between Z(3) vacua for parameter set 1 is very large compared 
to the parameter set 2 at any temperature greater than $\text T_0$. Fig.1 shows 
the barrier between different Z(3) vacua at a temperature $\text T= 400 MeV$ at 
the corresponding vacuum expectation value of Polyakov potential.
Here the parameters are such that the vacuum expectation value of 
Polyakov potential at T = 400 MeV for parameter set 1 and set 2
are 0.92 and 0.81 respectively. The energy density is almost equal
for different Z(3) vacua for both cases.

In the deconfined phase, due to breaking of Z(3) symmetry, one gets domain walls or interfaces 
that interpolate between different Z(3) vacua.
We can also see that Z(3) interface profile will be different for 
the Polyakov loop potential described by set 1 and set 2.     
The Z(3) interface arises as one goes from one Z(3) vacuum to another
Z(3) vacuum. These interfaces have been well studied in SU(3) lattice
pure gauge theory \cite{kaja}. The interface solution for the Polyakov loop
potential as a time independent solution is given in \cite{layek}. Here, 
we also use the same energy minimization technique as used in \cite{layek}
to get the interface profile.  To  determine  the interface
profile one needs to consider the profile of Polyakov loop in one dimension ( say along
z). We fix the values of $\Phi$ at the two boundaries of the one-dimensional lattice as
the values of Polyakov loop corresponding to two distinct minima (two Z(3) degenerate 
minima) of the Polyakov potential. Field configuration is then fluctuated at 
each lattice point, while fixing the boundary points, and energy is minimized.
The configuration with the lowest value of energy is accepted (when 
the energy almost settles down to a definite value)  as  the  correct  profile  of  
the  interface. We refer to \cite{layek} for the details
of the energy minimization technique. For the Polyakov loop potential with 
parameter set 1, the minimum energy configuration is such that 
the Polyakov order parameter goes very close to $\Phi=0$, since there 
is huge barrier between different Z(3) vacua. This is also seen in 
lattice gauge theory \cite{kaja}. However, for the potential with parameter 
set 2, the barrier between different Z(3) vacua is even smaller than 
the central bump near $\Phi$=0. So for the potential set 2, the minimum 
energy configuration going from one
vacuum to another vacuum never goes close to $\Phi=0$, rather it goes 
through a path where modulus of $\Phi$ remains almost constant and $\theta$
continuously changing. The profile of the interface in the complex plane 
of Polyakov loop order parameter at a temperature of 500 MeV ($\sim 2\text T_0$) for 
both parameter sets is shown in Fig.2.            

\begin{figure*}[!hpt]
\begin{center}
\leavevmode
\epsfysize=6truecm \vbox{\epsfbox{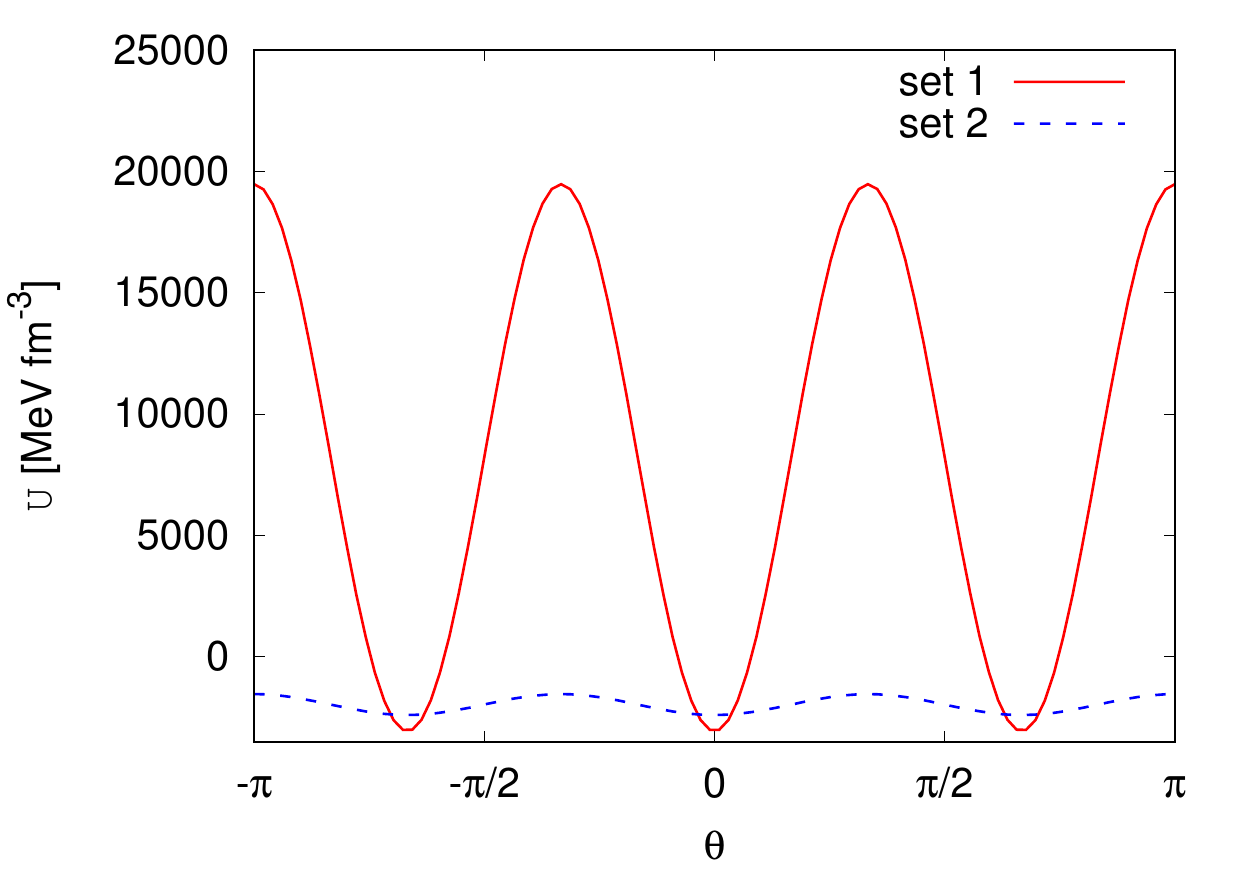}}
\end{center}
\caption{}{The Z(3) structure of the vacuum in the plot of Polyakov 
potential for both parameter sets as a function of $\theta$ at a 
temperature 400 MeV.}
\label{Fig.1}
\end{figure*}

\begin{figure*}[!hpt]
\begin{center}
\leavevmode
\epsfysize=6truecm \vbox{\epsfbox{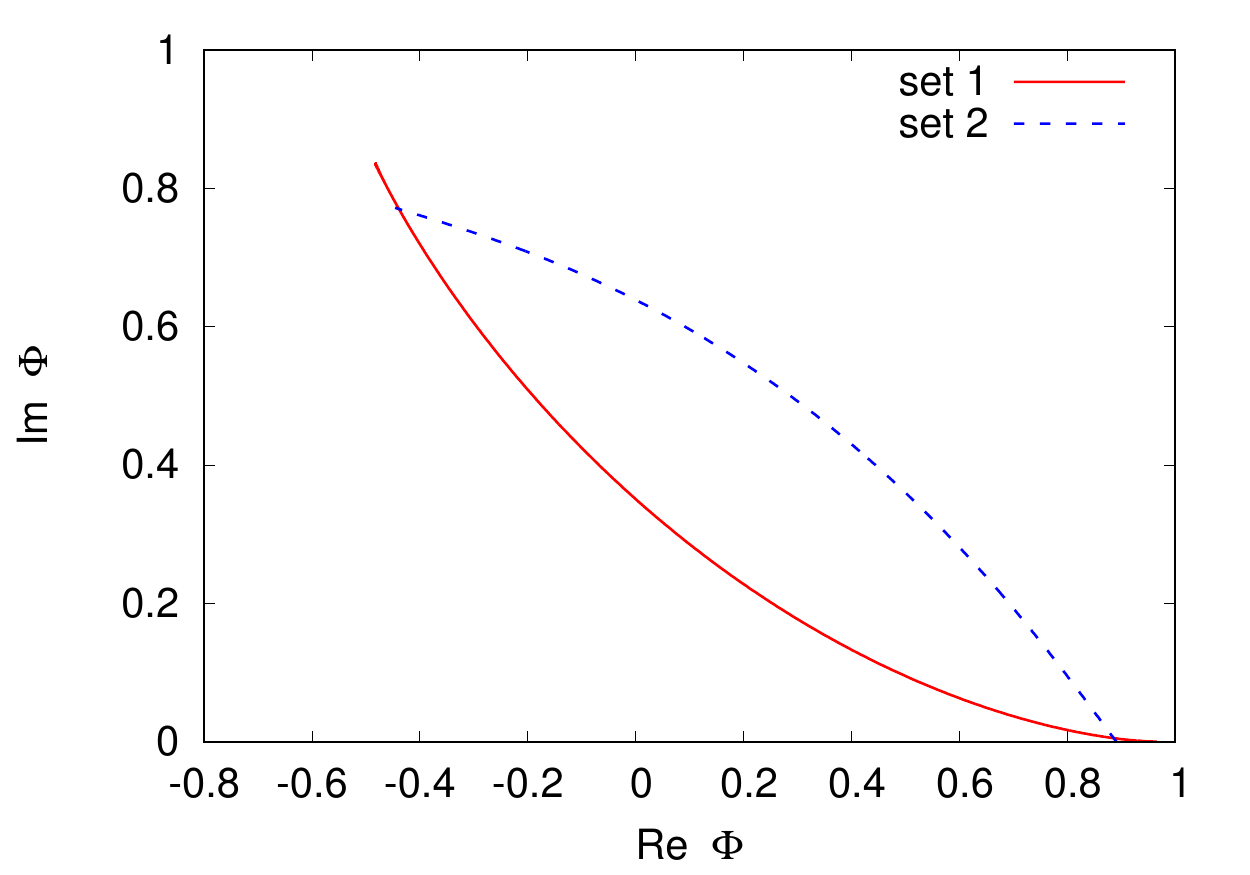}}
\end{center}
\caption{}{The path of the Polyakov order parameter $\Phi$ in the 
complex plane when crossing Z(3) interface at a temperature of $500 MeV$.
This path corresponds to a minimum energy configuration as one goes from
one vacuum to another Z(3) vacuum.} 
\label{Fig.2}
\end{figure*}

There is another possible parameterization used in the literature for the
Polyakov loop potential called logarithmic parameterization given by \cite{rose}

\begin{eqnarray}
&&\hspace{-1.4cm}\frac{{\cal U}_{log}(\Phi,\bar\Phi)}{\text T^4}=-\frac{1}{2}
A(\text T)\bar\Phi\Phi+B(\text T)\ln\left[1-6(\bar\Phi\Phi)+4(\Phi^3+(\bar\Phi)^3)-3
(\bar\Phi\Phi)^2\right].
\end{eqnarray}
Here, the temperature dependent coefficients are given by

\begin{equation}
A(\text T) = a_0  + a_1 \left(\frac{\text T_0}{\text T}\right) + a_2
\left(\frac{\text T_0}{\text T}\right)^2,
\end{equation}
and,

\begin{equation}
B(\text T) = b_3 \left(\frac{\text T_0}{\text T}\right)^3.
\end{equation}
The different parameters here are $a_0 = 3.51$, $a_1 = -2.47$, $a_2 = 15.2$, $b_3 = -1.75$

In this parametrisation, the effective potential between different Z(3) vacua 
is not well defined  at a temperature greater than $\text T_0$
as the argument of the logarithm in Eq.(8) can become negative.
Further, the Z(3) structure of 
the vacuum at the vacuum expectation value as a function of $\theta$ is not well defined for this
logarithmic potential.
Since we are interested in the 
details of Z(3) parameterization structure, we do not consider this parametrisation here.
 
In general, the Polyakov loop potential with parameter set 2
is  more commonly used in the context of PQM model because one is interested in 
the study of thermodynamic equilibrium properties of the system along the 
true vacuum\cite{scfer,herold,mintz}. On the other hand, since, we are interested in 
Z(3) metastability and the evolution of Z(3) domains with dynamical quarks, we use 
the Polyakov loop potential with parameter set 1 throughout this paper. 
We will also see later in the paper that including quark contribution to 
the Polyakov loop potential washes away Z(3) metastability structure for parameter set 2.
This is primarily because the barrier between different Z(3) vacua is very small for the 
parameter set 2. Since the barrier between different Z(3) vacua is very large for parameter 
set 1, the quark effect does not wash away the Z(3) metastability structure.

\subsection{Quark-meson coupling to Polyakov loop}    
 
Chiral symmetry of QCD is an  important symmetry to understand low 
energy hadronic properties \cite{koch}. There are different 
phenomenological models like Nambu-Jona-Lasinio (NJL) and Quark-Meson
(QM) model which are based on this chiral symmetry of QCD. By combining the 
Polyakov loop model with the QM model, both the confining and chiral 
properties of QCD are included.  
 
The Lagrangian of the linear QM model for N$_f$ = 2 light quarks 
$q = (u,d)$ and N = 3 color degrees of freedom coupled minimally to
a spatially constant temporal background gauge field is given by

\begin{eqnarray} 
\label{potomega}
  {\cal L} &=& \bar{q} \,[(i\gamma^{\mu}\partial_{\mu}-ig_{s}\gamma^{0}A_{0}) 
- g (\sigma + i \gamma_5 \vec \tau \vec \pi )]\,q
  +\frac 1 2 (\partial_\mu \sigma)^2+ \frac{ 1}{2}
  (\partial_\mu \vec \pi)^2
  \nonumber \\
  && \qquad - U(\sigma, \vec \pi )  -{\cal U}(\Phi,\bar\Phi)
\end{eqnarray}
where the linear sigma model potential reads 
\begin{eqnarray} 
\label{pots}
U(\sigma, \vec \pi ) &=& \frac \lambda 4 (\sigma^2+\vec \pi^2 -v^2)^2
-c\sigma
\end{eqnarray}
The parameters in Eq.(\ref{pots}) are chosen such that chiral symmetry 
is spontaneously broken in vacuum, where $\langle\sigma\rangle = f_\pi 
= 93$ MeV, and here $f_\pi$ is the pion decay constant. Since pions are 
pseudoscalar in character, the expectation values vanish 
$\langle {\vec \pi}\rangle =0$. The explicit symmetry breaking term is
$c \sim 1.77\times 10^6$ MeV$^3$ which produces a pion mass of 138 MeV.
The quartic coupling $\lambda$ is given by sigma mass $m_{\sigma}$
by the relation $\lambda= (m_\sigma^2 -m_\pi^2)/ {2f_\pi^2}$.In the present 
calculations, we take $m_\sigma=600 $ MeV leading to $\lambda\simeq 20$. 
The parameter $v^2$ is found by minimizing the potential in radial 
direction $v^2 = {\sigma}^2-c/(\lambda \sigma)$. Finally, 
the Yukawa coupling constant $g = 3.2$ which is fixed to produce a 
constituent quark mass of 300 MeV in the vacuum $m_q = gf_\pi$. 
 
The partition function $\cal Z$ is written as a path integral
over quarks, anti-quarks, mesons and the temporal component of the
gauge field. We shall adopt here a mean field approximation of
replacing the meson and the Polyakov loop fields by their vacuum expectation values.
This amounts to neglecting both the quantum and thermal fluctuations of all the fields other than
the quark and anti quark fields. Integrating over the quark degrees of freedom, one can 
get the thermodynamic potential.
The effective thermodynamic potential is determined as the logarithm of
the partition function.

\begin{equation}
\label{veff}
V_{eff} =-\frac{\text T}{\text V} \ln {\cal Z} ={\cal U} (\Phi,\bar\Phi) +
  U(\sigma) + \Omega_{\bar qq}{(\Phi,\bar\Phi,\sigma)}
\end{equation}

where

\begin{eqnarray} 
\label{omegaqq}
\Omega_{\bar qq} &=& -2\text N_f \text T\int \frac{d^3p}{(2\pi)^3}
\ln \left[1 + 3 (\Phi + \bar \Phi e^{-(E_p-\mu)/\text T})e^{-(E_p-\mu)/\text T}+
e^{-3(E_p-\mu)/\text T}\right] 
\nonumber \\ && \qquad +\ln \left[1 +3 (\bar \Phi + \Phi
e^{-(E_p+\mu)/\text T})e^{-(E_p+\mu)/\text T} + e^{-3(E_p+\mu)/\text T}\right].
\end{eqnarray}
In the above, $E_{p}$ is the quark or anti-quark quasi particle energy given by
 $ E_p = \sqrt{|\vec p|^2 + m_q^2}$
with the constituent quark mass $m_{q}=g\sigma$. Clearly in Eq.(\ref{omegaqq}) above 
we have written $\sigma=\langle\sigma\rangle$, $\Phi=\langle\Phi\rangle$, 
$\bar\Phi=\langle\Phi\rangle$ and have taken pion field as having vanishing 
vacuum expectation value. Further, in Eq.(\ref{omegaqq}), we have omitted
a zero temperature and density contribution in $\Omega_{\bar qq}$ that can 
be absorbed partly into parameter $\lambda$ and a logarithmic term depending 
upon the renormalization scale and effective quark mass. However, the qualitative 
features of the phase diagram remains unchanged as long as the mass of the sigma 
meson is not too high \cite{mintz,wagner}. We therefore proceed our 
analysis without the vacuum fluctuation terms for the fermions in $\Omega_{\bar qq}$.

\section{Z(3) METASTABILITY IN PQM MODEL}

In pure SU(3) gauge theory, the deconfined phase exists in three
degenerate states and these three states are related to each other 
via Z(3) rotation. However, inclusion of dynamical quarks, breaks this
Z(3) symmetry due to anti periodic boundary conditions on fermions.
Thus, the Z(3) symmetry is explicitly broken giving rise to one true vacuum
along $\theta = 0$ and two metastable vacua along $\theta=2\pi/3$ and
4$\pi/3$. In this section, we will discuss about these Z(3) metastable 
states in the context of PQM model where the effect of quarks are included 
in terms of QM model.

To study the dynamics of phase transition within PQM model,it is convenient
we write down the effective thermodynamic potential Eq.(\ref{veff})
in terms of the real and imaginary parts of Polyakov order parameter as in Ref. \cite{mintz}. 
Defining $\alpha=(\Phi+\bar\Phi)/2=|\Phi|\cos\theta$
and $\beta=(\Phi-\bar\Phi)/(2i)=|\Phi|\sin\theta$,
the Polyakov potential for parameter set 1 i.e Eq.(\ref{set1v}) and 
set 2 i.e Eq.(\ref{set2v}) can be rewritten as functions of real variable $\alpha$ 
and $\beta$  respectively as

\begin{subequations}
\begin{eqnarray}
\label{potalpha}
 \frac{{\cal U}(\alpha,\beta)}{\text T^4} =
 \left(-\frac{b_2}2(\alpha^2+\beta^2) - \frac{b_3}3(\alpha^3 -
 3\alpha\beta^2) + \frac{1}{4}(\alpha^2 + \beta^2)^2\right)b_4\\
\frac{{\cal U}(\alpha,\beta)}{\text T^4} =
 \left(-\frac{b_2}2(\alpha^2+\beta^2) - \frac{b_3}3(\alpha^3 -
 3\alpha\beta^2) + \frac{b_4}{16}(\alpha^2 + \beta^2)^2\right)
\end{eqnarray}
\end{subequations}

Similarly, defining
\begin{equation}
 x_+ \equiv {1 + 3(\Phi + \bar \Phi e^{-(E-\mu)/\text T} )e^{-(E-\mu)/\text T} +e^{-3(E-\mu)/\text T}} ,
\end{equation}
and

\begin{equation}
 x_- \equiv 1 + 3(\bar \Phi + \Phi e^{-(E+\mu)/\text T} )e^{-(E+\mu)/\text T} +
 e^{-3(E+\mu)/\text T},
\end{equation}

\begin{equation}
 x_+x_- = R+iI,
\end{equation}

the contribution $\Omega_{q\bar q}$ from the quarks at finite temperature and chemical
potential can be written as 

\begin{equation}
 \Omega_{q\bar q} = -2\text N_f\text T\int\frac{d^3p}{(2\pi)^3}\log[x_+x_-].
\end{equation}
 It can be seen that the argument of the logarithm is complex and $x_+x_-=R+iI$
where

\begin{eqnarray}\label{eq:real_part_arglog_qqbar}
 R \equiv && 1 + e^{-3(E-\mu)/\text T} + e^{-3(E+\mu)/\text T} + e^{-6E/\text T} +
 \nonumber\\ &&  6\alpha e^{-E/\text T}\left[\cosh\left(\frac{\mu}{ \text T}\right) +
   e^{-E/\text T}\cosh\left(\frac{2\mu}{\text T}\right)\right] + \nonumber\\ && 
 6\alpha e^{-4E/\text T}\left[\cosh\left(\frac{2\mu}{\text T}\right) +
   e^{-E/\text T}\cosh\left(\frac{\mu}{ \text T}\right)\right] + \nonumber\\ && 
 9(\alpha^2+\beta^2)(1+e^{-2E/\text T})e^{-2E/\text T} + \nonumber\\ && 
 18(\alpha^2-\beta^2)e^{-3E/\text T}\cosh\left(\frac{\mu }{\text T}\right)\;,
\end{eqnarray}

and

\begin{eqnarray}\label{eq:imaginary_part_arglog_qqbar}
 I \equiv && 6\beta e^{-E/\text T}\left[\sinh\left(\frac{\mu}{ \text T}\right) -
   e^{-E/\text T}\sinh\left(\frac{2\mu}{\text T}\right)\right] + \nonumber\\ && 
 6\beta e^{-4E/\text T}\left[e^{-E/\text T}\sinh\left(\frac{\mu}{ \text T}\right) -
   \sinh\left(\frac{2\mu}{\text T}\right)\right] - \nonumber\\ && 
 36\alpha\beta\sinh\left(\frac{\mu}{ \text T}\right)e^{-3E/\text T}.
\end{eqnarray}
In principle, one can write $R+iI=\rho e^{i\delta}$
where 

\begin{equation}
 \rho \equiv\sqrt{R^2 +
   I^2}\;\;\;\mbox{and}\;\;\;\delta \equiv\arctan\left(\frac IR\right)\;,
\end{equation}
Thus the  potential has an imaginary part and this is the 
manifestation of fermion sign problem in the context of PQM model
\cite{mintz,rosner}.

\begin{equation}
 \Omega_{q\bar q} = \Omega_{q\bar q}^R + i\,\Omega_{q\bar q}^I \;,
\label{omega}
\end{equation}

with

\begin{equation}
\label{omega_r}
 \Omega_{q\bar q}^R\equiv-2\text N_f\text T\int\frac{d^3p}{(2\pi)^3}\ln[\rho] \;,
\end{equation}
and

\begin{equation}
\label{omega_im}
 \Omega_{q\bar q}^I\equiv -2\text N_f\text T\int\frac{d^3p}{(2\pi)^3}\delta
\end{equation}

\subsection{ Z(3) metastable states at zero chemical potential}
 
The imaginary part of the potential i.e Eq.(\ref{omega_im})  vanishes at 
$\mu = 0$ and the effective potential becomes real. There is no fermion 
sign problem at zero chemical potential. We will first consider 
the case with $\mu = 0$. For two flavors, we take the value of $\text T_0$ as $210$ MeV
\cite{scfer}. When the real part of Polyakov loop $\alpha$ is positive, 
 $\beta$=0 (i.e along $\theta$ = 0) and appropriate $\sigma$ value corresponding to 
the temperature, $\Omega_{q\bar q }^R$ in Eq.(\ref{omega_r}) becomes negative.
So, the total effective potential $V_{eff}$ along $\theta$ =0 is a global minimum and it 
becomes the true vacuum. However, along other two Z(3) vacua, when $\alpha$ is negative, 
$\Omega_{q \bar q }^R$ in Eq.(\ref{omega_r}) is positive. Hence the total effective 
potential $V_{eff}$ can have metastable vacua
along these directions. For the two different Polyakov loop potential set 1 
and set 2, the effective potential at the vacuum expectation value of Polyakov loop 
potential as a function of $\theta$ at temperature 400 MeV is shown in Fig.3.

\begin{figure*}[!hpt]
\begin{center}
\leavevmode
\epsfysize=6truecm \vbox{\epsfbox{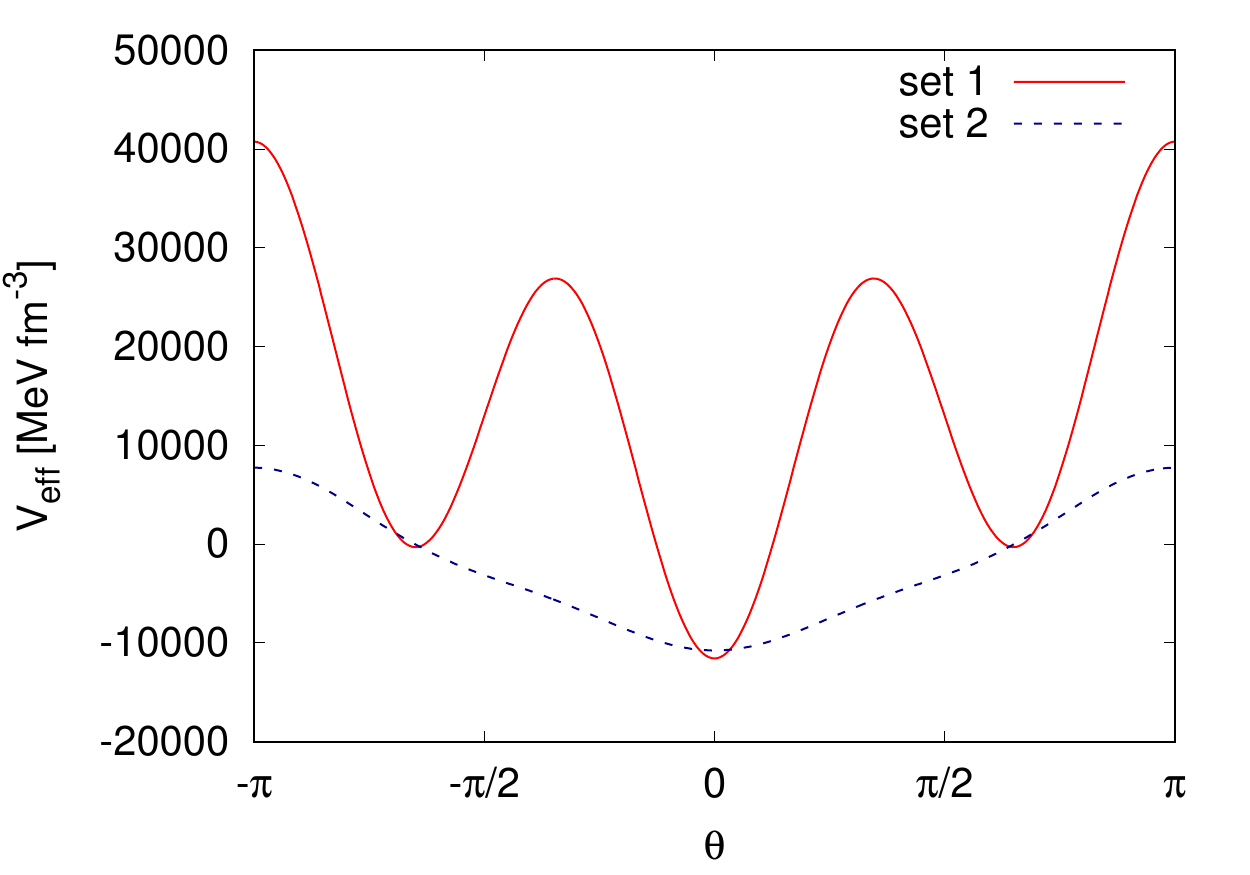}}
\end{center}
\caption{}{ Effective potential as a function of phase of Polyakov loop for $\text T=400$ MeV 
and $\mu=0$. Set 1 corresponds to
the Polyakov potential of Ref.\cite{psrsk2} while Set 2 corresponds to Ref \cite{scfer}.
}
\label{Fig.3}
\end{figure*}

From Fig.3, it is clear that there are no metastable vacua for Polyakov 
loop potential set 2 since the barrier between different Z(3) vacua is 
very small for SU(3) pure gauge theory as shown in Fig.1 and strong 
explicit symmetry breaking effect due to quarks completely wash away 
the metastable structure. For parameter set 2, the 
potential is tilted all the way to $\theta$ = 0 vacuum at all 
temperatures greater than critical temperature. But, this is not the 
case for the parameter set 1. One can clearly see the metastable and true vacua 
in PQM model for Polyakov loop potential corresponding to set 1.  
Since we are interested in Z(3) vacuum structure, we consider 
the Polyakov loop potential set 1 throughout the paper. 

By minimizing the effective thermodynamic potential $V_{eff}$ with
respect to $\alpha$, $\beta$ and $\sigma$, 

\begin{equation}
\label{eqeom}
\frac{\partial V_{eff}}{\partial \alpha}=\frac{\partial V_{eff}}{\partial
  \beta} = \frac{\partial V_{eff}}{\partial \sigma}= 0
\end{equation}
we get the global minimum at finite value of $\alpha$ and $\sigma$ and 
$\beta = 0$ (along $\theta = 0$ direction). A comment regarding the above
may be relevant. The condition of Eq.(\ref{eqeom}) is a necessary condition for a minimum
and not a sufficient condition. We have verified that the solution is indeed 
a minimum and does not correspond to a maximum or a saddle point.
The variation of modulus of the Polyakov loop order parameter 
$\left|\Phi\right|=\sqrt{(\alpha^2+\beta^2)}$
and $\sigma$ with respect to temperature
is shown in Fig.4. Here, we can see a chiral crossover is found at
a temperature $\text T= 180$ MeV as also seen in \cite{scfer}. We also find that 
$\Phi$ reaches a higher value 1.11 as temperature increases \cite{scfer}.  

\begin{figure*}[!hpt]
\begin{center}
\leavevmode
\epsfysize=6truecm \vbox{\epsfbox{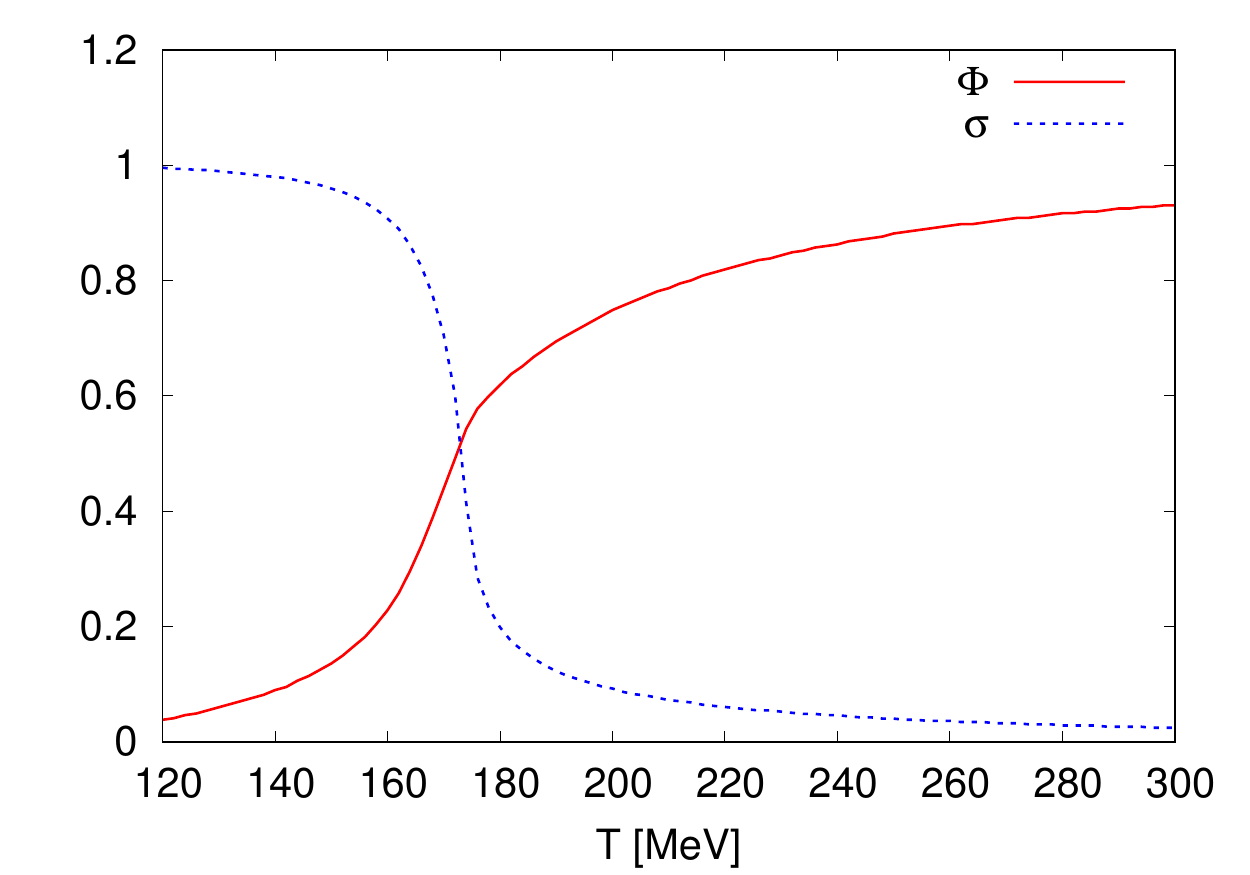}}
\end{center}
\caption{}{The normalized chiral condensate $\sigma$ and the Polyakov 
loop $\Phi$ as a function of temperature.}
\label{Fig.4}
\end{figure*}

Now using the same $\sigma$ value at a given 
temperature that minimizes the effective potential $V_{eff}$, we 
scan all values of ($\alpha$, $\beta$) corresponding to the second and the third 
quadrant of the Polyakov 
loop order parameter to find local minima corresponding to Z(3) metastable vacua.
 We ascribe the state as metastable when the corresponding pressure is positive and is lower than
the pressure at $\theta=0$. We would like to note here that, this procedure
is only an approximate method to get a local minimum. In principle, there could 
be a different pair of $(\sigma,\Phi)$ which can still be metastable with a 
value of $\sigma$ other than the value of $\sigma$ at the
minimum with $\theta=0$.
We do not see any Z(3) metastable vacua up to a temperature of 310 MeV,
i.e the free energy density is still higher than the free energy density in the 
confined phase. At a temperature of 310 MeV and beyond, Z(3) metastable 
vacua start appearing. However, there is a large difference of free energy density
($\epsilon \sim 4.0$ GeV fm$^{-3}$) between true and Z(3) metastable vacua at this
temperature. This energy difference between metastable and true vacua
increases as temperature increases as shown in Fig.5. It has been already shown that the 
free energy density difference between true and metastable vacua varies as $\frac{2}{3}
\pi^2\text T^4\frac{\text N_{l}}{\text N^3}(\text N^2-2)$ using perturbative calculation \cite{dixit}. Here,
$\text N_{l}=2$ is the number of massless fermions and $\text N=3$ is the no. of colors.  
In Fig.5,  we have fitted the plot by using the same function $f(\text T)=\frac{2}{3}
\pi^2\text T^4\frac{\text N_{l}}{\text N^3}(\text N^2-2)$. This function fits with PQM model
extraordinarily well beyond temperature 550 MeV( as shown in inset). However, near 
temperature $\sim 310$ MeV when Z(3) metastable vacua start appearing, this function 
does not fit well with data. The difference in the free energy density is 
$\sim 0.3$ GeV fm$^{-3}$ near temperature 310 MeV. This is mainly because the 
functional dependence given above comes from perturbative calculations which 
is valid at higher temperatures.

\begin{figure*}[!hpt]
\begin{center}
\leavevmode
\epsfysize=6truecm \vbox{\epsfbox{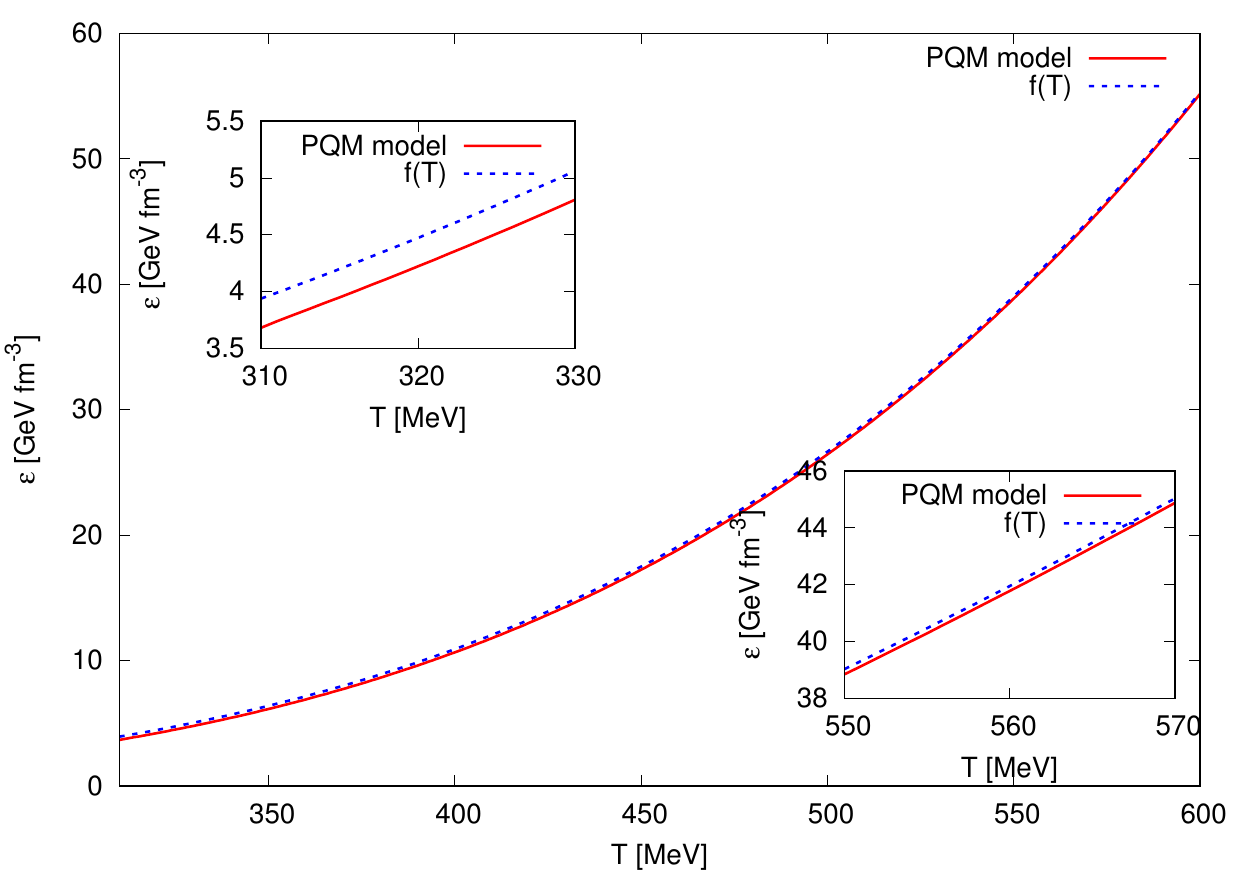}}
\end{center}
\caption{}{Free Energy density difference between true and metastable vacua as 
a function of temperature}

\label{Fig.5}
\end{figure*}

Further, there is a shift in the phase of the Polyakov loop
the metastable vacua of the order of $5^{\circ}$ (metastable vacua
appear at  angles $115^{\circ}$ and $245^{\circ}$). The magnitude
of $\Phi$ along the metastable vacua is also smaller than the
magnitude along the true vacuum.  We show contour plots of the 
effective potential $V_{eff}$ as a function of 
real and imaginary parts of the Polyakov loop at different temperatures 
in Fig.6 . Here the value of $\sigma$ is taken  as the value 
which minimizes the effective potential $V_{eff}$ at the corresponding
 temperature. The maximum value 
of $V_{eff}$ is set as 6.3 GeV fm$^{-3}$ in all figures to show the 
true, metastable vacua and the barrier between them distinctively.   
Fig.6a shows the contour plot of the effective potential at temperature 300 MeV
and we can clearly see that there is no Z(3) metastable vacua at this temperature
since the energy density of metastable vacuum is still positive ( i.e larger than 
the energy density in the confined phase). Fig.6b represents the effective potential at 
temperature 400 MeV and Z(3) metastable vacua and true vacuum are clearly seen 
at this temperature.
 
\begin{figure*}[!hpt]
\begin{center}
\leavevmode
\epsfysize=6truecm \vbox{\epsfbox{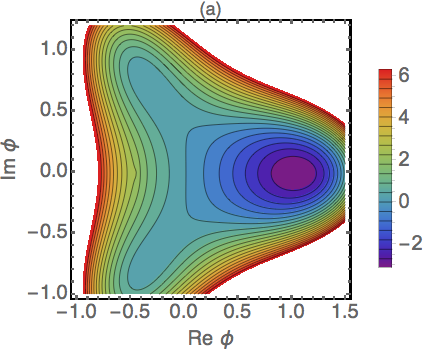}}
\epsfysize=6truecm \vbox{\epsfbox{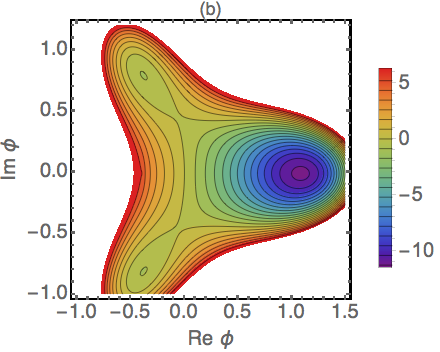}}
\end{center}

\caption{}{Here (a) and (b)represent the contour plot of 
effective potential at different temperatures 300 MeV and 
400 MeV respectively. Here the legend bar represents energy density 
in GeV fm$^{-3}$}
\label{Fig.6}
\end{figure*}

Next, we  examine another case where explicit symmetry breaking 
effect due to quarks is included as a linear term in the Polyakov loop in 
the Polyakov potential given in Eq.(4)\cite{psrsk2}. We will compare here the explicit 
symmetry breaking effects for this case with PQM model.  
The Polyakov loop potential with explicit symmetry breaking term is given by \cite{psrsk2}

\begin{eqnarray}
\label{set1vb1}
&&\hspace{-1.4cm}\frac{{{\cal U}_e}(\Phi,\bar\Phi)}{\text T^4}=
\left(-b_{1}\left(|\Phi|\right)\cos\theta-\frac{b_2}{4}
\left(|\Phi|^2+|\bar\Phi|^2 \right)
-\frac{b_3}{6}(\Phi^3+\bar\Phi^3)+\frac{1}{16}
\left(|\Phi|^2+|\bar\Phi|^2\right)^2\right)*b_4
\end{eqnarray}

Here, $b_{1}$ is the explicit symmetry breaking term which takes values in the range 
from 0 to 0.12 as shown in Ref. \cite{psrsk2}. 
While for small $b_1$ ($b_1<0.026$), the weakly first order transition of pure gauge theory persists, the same becomes a cross over for
 larger values of $b_1$. For comparison  with PQM model where the transition is a cross over, we have taken a value $b_1=0.1$. Further,
to take into account the flavor effect, the parameter $b_4$ is multiplied by a 
factor $37/16$ to its pure gauge value.
In Ref.\cite{psrsk2}, it was shown that this explicit 
symmetry breaking linear term is sufficient to describe the dependence of the chiral symmetry 
restoration temperature on pion mass. 
For this potential given in Eq.(\ref{set1vb1}), Z(3) metastable vacua are there at all temperatures greater than the 
critical temperature and the phase shift of the metastable vacuum is negligible.

In Fig.7 we have plotted the explicit symmetry breaking potential ${\cal U}_e$ of 
Eq.(\ref{set1vb1}) for  $\theta=0$  corresponding to the stable vacuum and 
for $\theta=2\pi/3$ corresponding to metastable vacuum at a temperature 400 MeV 
as a function of the order parameter $|\Phi|$. 
These are shown by the red lines. In the same figure we have also plotted the 
$V_{eff}$ of Eq.(\ref{veff}) the corresponding curves for the stable ( $\theta=0$) as well as
the metastable ($\theta=115^{\circ}$) vacuum. This because, as mentioned earlier, 
the metastable vacuum for the PQM
case occurs with a shift of $5^{\circ}$.

It is clear from the figure that the free energy density difference between 
true and Z(3) metastable vacuum is very small in case of small explicit symmetry 
breaking term used in \cite{psrsk2} as
compared to PQM model.  Thus, PQM model shows strong explicit breaking effects due to
quarks.  Further, the energy density difference along metastable and true vacuum near
$\Phi=0$ is large in PQM model compared to the potential in Eq.(\ref{set1vb1}).
This will have important consequences for the phase transition kinetics for these two cases
as we shall see in the next section.  We will see 
later in the simulation that when the explicit symmetry breaking is 
small \cite{psrsk2} with $b_1=0.1$, Z(3) domains
are formed during the initial time of evolution after quenching the system to a higher 
temperature of 400 MeV. Subsequently the metastable domains collapse and true 
vacuum domains expand due to the difference in free energy between them. On the other hand,
 for PQM model, after the quench the whole system evolves into true vacuum 
rapidly. For the large explicit symmetry breaking in PQM 
model, we do not get any metastable domains in simulation. However, this is not the case with 
small explicit symmetry breaking effect used in \cite{psrsk2}.      
This will be discussed later in detail in the next section.

\begin{figure*}[!hpt]
\begin{center}
\leavevmode
\epsfysize=6truecm \vbox{\epsfbox{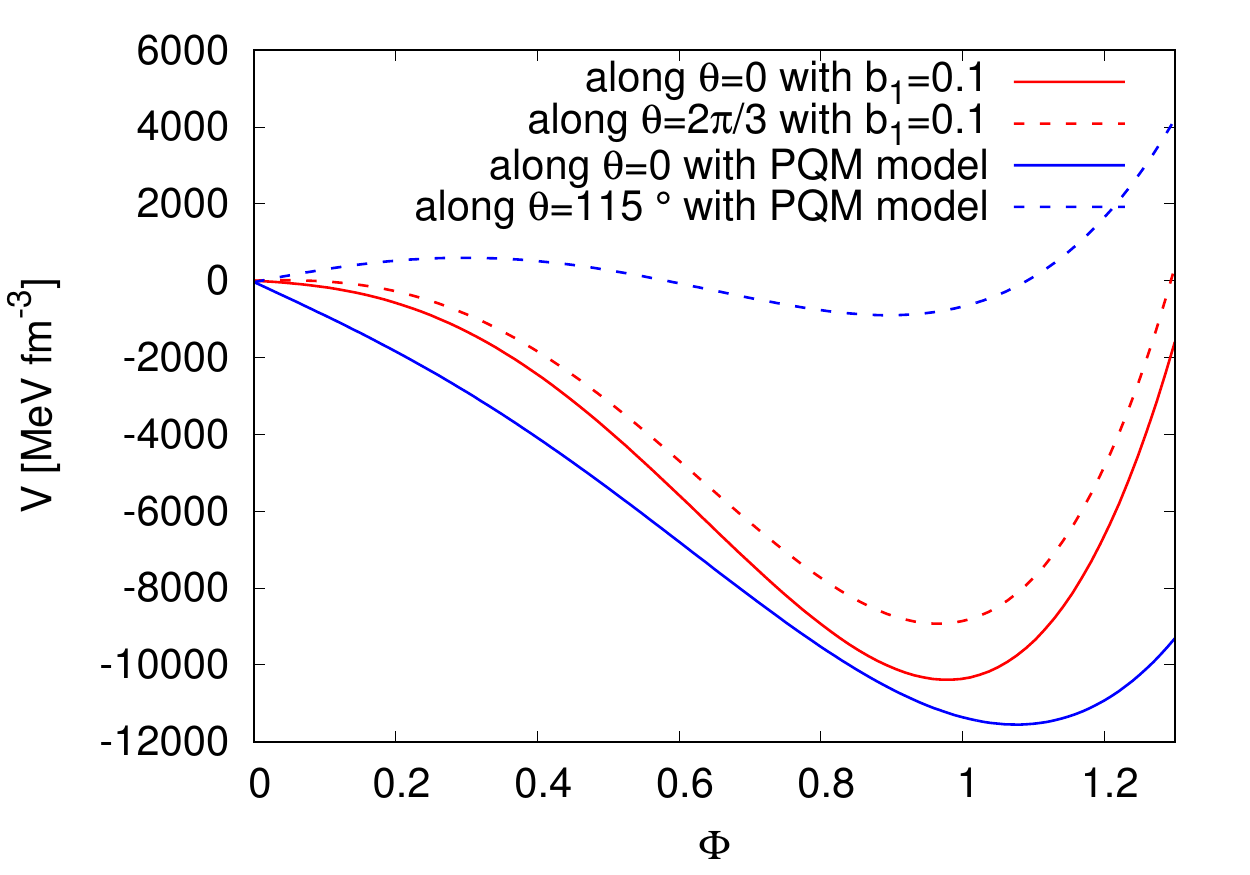}}
\end{center}
\caption{}{The Polyakov loop potential with explicit symmetry breaking 
term ${\cal U}_e$ (Eq.(\ref{set1vb1})) 
and the effective potential $V_{eff}$ of  PQM model (Eq.(\ref{veff}))
as a function of $\Phi$ along true and metastable
vacua at T = 400 MeV}
\label{Fig.7}
\end{figure*}

Here we would like to explore another interesting possibility of the dependence of
existence of metastable states on no. of flavors.
The variation of the effective potential on the   number of flavors 
was discussed in Ref.s \cite{dixit,monnai}. It has been shown in these references
that as the number of flavors 
increases, the free energy density of metastable vacua increases at any 
temperature. For $N_f \ge 3$,  no metastability 
at any temperature was seen. However, these are essentially perturbative calculations 
which are not applicable near $\text T_c$. 
In the PQM model discussed here,
the flavor dependence comes from the third term in the effective potential 
$V_{eff}$ (Eq.(\ref{veff})). The parameters of linear sigma model potential part 
in Eq.(\ref{veff}) are fixed to reproduce the results of chiral symmetry
restoration for $N_f$ = 2 . For simplicity, we take this 
linear sigma model part in Eq.(\ref{veff}) for different values of $N_f$. 
For the Polyakov loop part of the potential in Eq.(\ref{veff}), the parameter $\text T_0$
depends on the no. of flavors 
and chemical potential as \cite{scfer}, 

\begin{equation}
\label{tmu}
  \text T_0(\mu, N_f)=\text T_\tau e^{ -1/(\alpha_{0} b(\mu))}.
\end{equation}
Here, 
\begin{equation}
  b(\mu)=\frac{1}{6 \pi} (11 N_c-2 N_f)- b_{\mu}
\frac{\mu^2}{\text T_\tau^2}
\end{equation} 
where $\text T_{\tau}$=1.770 GeV, $\alpha_{0}$=0.304 and $b_{\mu}=16N_{f}/\pi$.
The flavor dependence of the effective potential normalized by $\text T^4$ is shown in Fig. 8
for $\mu=0$ and $\text T= 400$ MeV. As may be noted from the Fig. 8, the absolute value of 
the free energy difference between the true vacuum ($\theta=0$) and the minima 
at nonzero $\theta$ increases with the flavor number. It may also be noted that 
beyond $N_f\ge 3$, the metastable minima become unphysical as the pressure 
become negative. Such an observation was also noted 
in Ref.s \cite{dixit,monnai} within a perturbative approach.
For higher temperatures also we did not find any metastable state ( up to T=600 MeV).
  
\begin{figure*}[!hpt]
\begin{center}
\leavevmode
\epsfysize=6truecm \vbox{\epsfbox{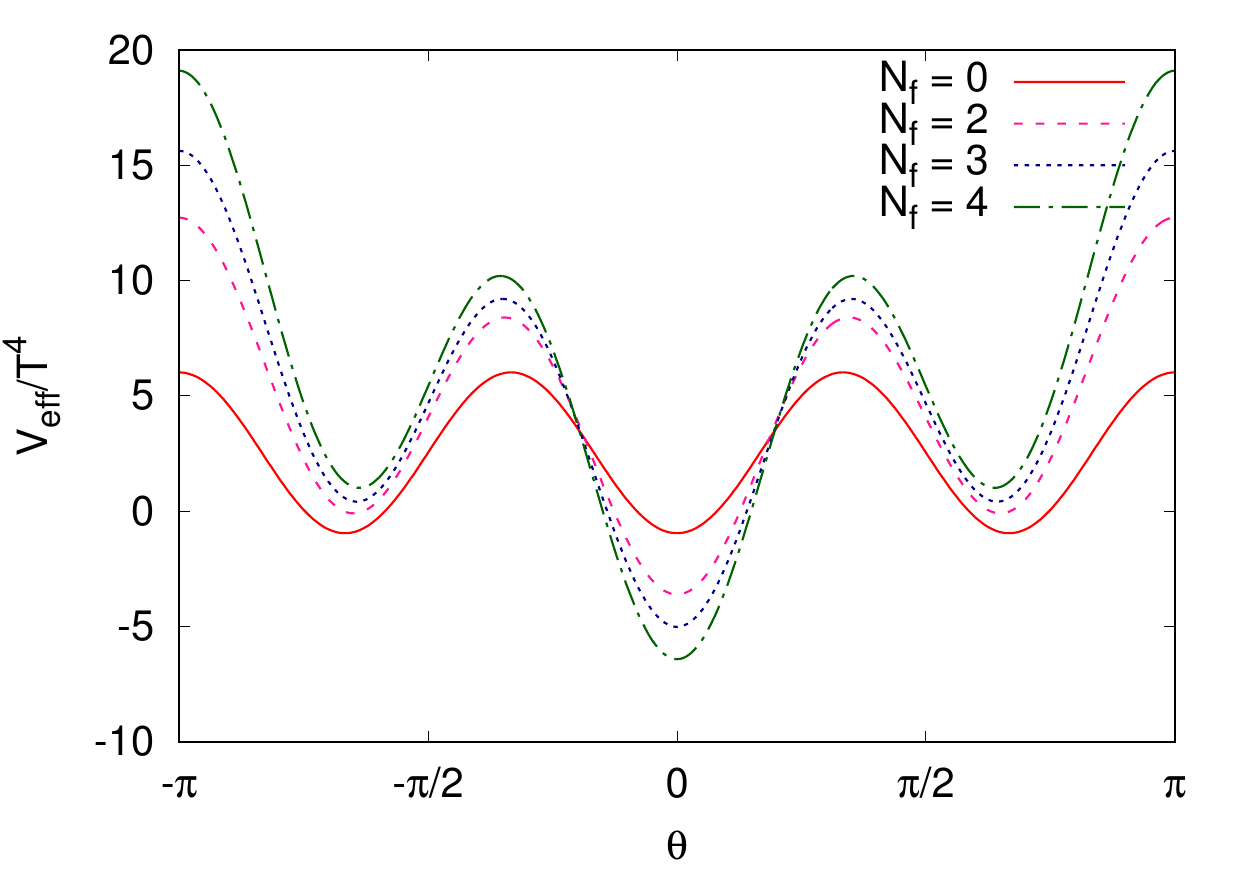}}
\end{center}
\caption{}{$V_{eff}$ as a function of $\theta$ for different no. of flavors
at T = 400 MeV}
\label{Fig.8}
\end{figure*}

\subsection{Meta stable states at finite chemical potential}

From Eq.(\ref{omega_im}), it is clear that there is a non vanishing contribution from 
imaginary part of $\Omega_{\bar qq}$ at finite chemical potential \cite{mintz}. However, 
as argued in Ref.\cite{mintz}, we neglect this part in our calculation at finite 
temperature so that the effective potential is real where a 
minimization procedure can be applied.
The critical temperature decreases as chemical potential increases 
as seen in \cite{scfer}. The variation of both the order parameters at 
different chemical potentials with respect to temperature is 
the same as seen in \cite{scfer}. 
The free energy density of true vacuum at finite chemical potential is more negative 
than the  same at zero chemical potential at any temperature greater
than the critical temperature. But, the energy density of Z(3) metastable vacua 
at finite chemical potential increase compared to zero chemical potential.
Hence, the threshold temperature for metastability  increases  with the chemical potential.
The temperature $\text T_m$ beyond which Z(3) metastable vacuum 
arise is shown as a function of chemical potential in Fig.9

\begin{figure*}[!hpt]
\begin{center}
\leavevmode
\epsfysize=6truecm \vbox{\epsfbox{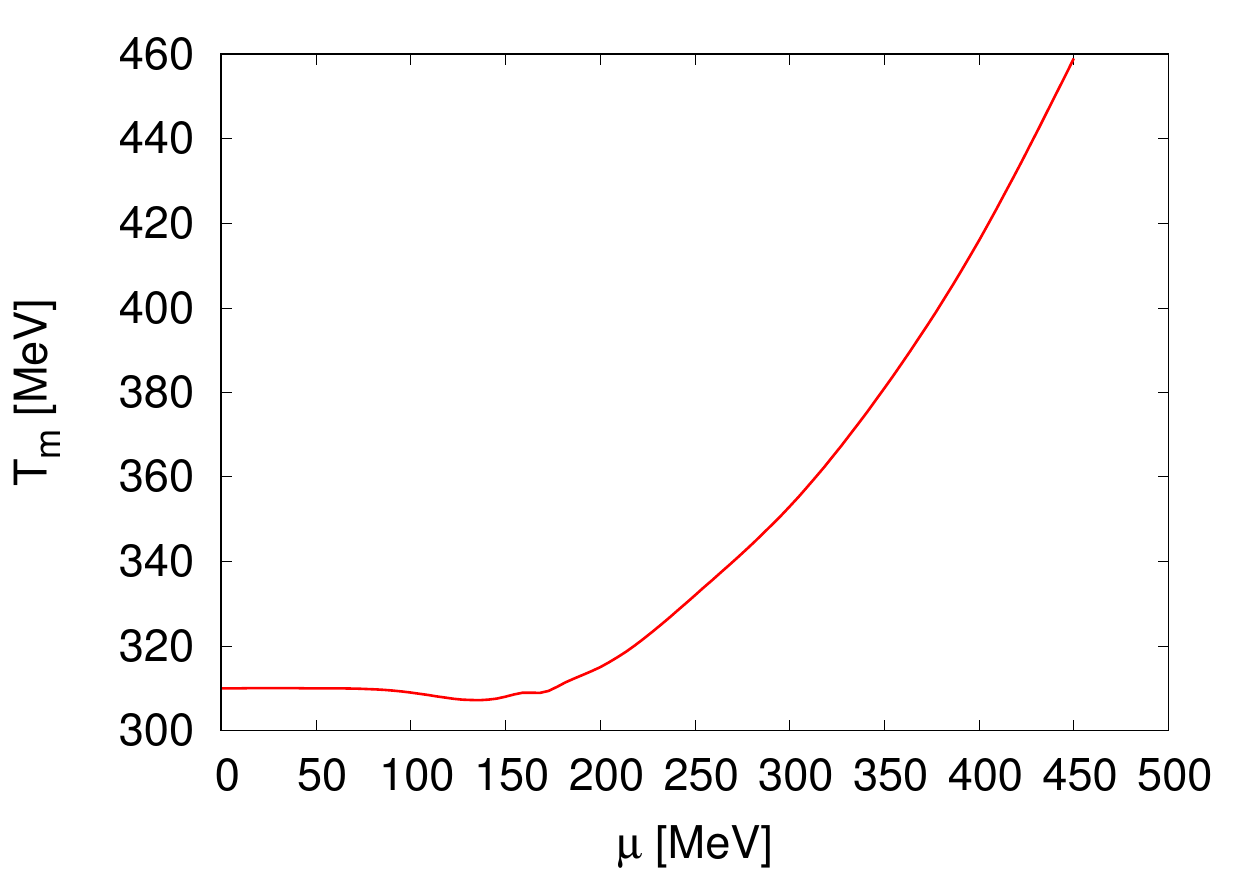}}
\end{center}
\caption{}{The temperature $\text T_m$ beyond which Z(3) metastable vacuum occurs 
as a function chemical potential.}
\label{Fig.9}
\end{figure*}

Here, we have taken the  temperature parameter
$\text T_{0}$ for the Polyakov loop potential  to have a  $\mu$ dependence  as  given  
in Eq.(\ref{tmu}). From Fig.9, one can see that the metastability temperature 
remains almost constant ($\text T_{m}= 310$ MeV) up to $\mu= 170$ MeV and it increases
beyond $\mu= 170$ MeV. Generally, the metastability temperature should increase with 
increase in chemical potential since we are adding more dynamical quark degrees
of freedom by increasing the chemical potential, however, the cause for this sharp
increase of metastability temperature beyond $\mu= 170$ MeV is not very clear. 
We expect that there must be drastic change in phase transition dynamics beyond 
this chemical potential. It has been already observed that the critical point exists
at $(\text T_{c},\mu_{c})$=(150 MeV,168 MeV) \cite{scfer}. When the chemical potential is less
than the critical chemical potential, it is a crossover in QCD phase diagram. Beyond
critical chemical potential, there is a first order phase transition between confined and 
deconfined phase where the dynamics of the phase transition are very different 
than the crossover. So, the temperature dependence of metastable states at 
higher chemical potential might tell about the phase transition dynamics and 
hints towards the existence of a critical point. These metastable states have 
a phase of $115 ^\circ$ as already seen 
for zero chemical potential.

\section{PHASE TRANSITION KINETICS AND NUMERICAL SIMULATION}

The thermalization time scale at RHIC is very small ($\sim 0.2$ fm), in this 
short time scale an equilibrium dynamics of the transition from confined phase to
deconfined phase appears very unlikely. Hence in this work, we carry out a $2+1$ 
dimensional field theoretical simulation of the dynamics of confinement-deconfinement 
transition in a quench as in Ref. \cite{ranj}. Here we use the 
framework of Bjorken's boost invariant 
longitudinal expansion model \cite{bjorken} for the central rapidity region 
in relativistic heavy ion collisions. To model the quench, we take the initial field 
configuration to constitute a small patch around $\Phi = 0$ 
which corresponds to confining vacuum  configuration near zero temperature.  We take the 
initial phase of $\Phi$ to vary randomly between 0 and $2\pi$ from one lattice site 
to the other, while the magnitude of $\Phi$ is taken appropriately to obtain the
Z(3) domain structure. In principle, the initial magnitude of $\Phi$ 
(i.e $\epsilon = 0.1\times vev$)
should be smaller than the vacuum expectation value (vev) of $\Phi$ at 
$\text T_{max} = 400$ MeV (i.e maximum temperature obtained at LHC).  

  This initial field configuration, which represents the equilibrium
field configuration of a system with $\text T << \text T_c$, is evolved using
the effective potential with $\text T = \text T_{max} > \text T_c$. This represents the
transition dynamics of a quench. Here, we will compare the two different cases:\\
\begin{itemize}
\item  Polyakov potential with explicit symmetry breaking term as in Eq.(\ref{set1vb1}).\\
\item PQM model effective potential as given in Eq.(\ref{veff}).
\end{itemize}

First we will consider the situation of Polyakov loop potential with explicit 
symmetry breaking term as given in Eq.(\ref{set1vb1}). The kinetic energy term for the Polyakov
loop field is represented by $\frac{\text N}{g^2}{|\partial_{\mu}\Phi|}^2\text T^2$. The field 
configuration of Polyakov loop field is evolved by the time dependent equation 
of motion in the Minkowski space as appropriate for Bjorken’s longitudinal 
scaling model \cite{rndrp}

\begin{equation}
\label{eomphi}
  \frac{\partial^{2}{\Phi}_j}{\partial\tau^{2}} + \frac{1}{\tau} 
\frac{\partial{\Phi}_j}{\partial \tau}  -\frac{\partial ^{2}{\Phi}_j}
{\partial x^{2}}   
-\frac{\partial^{2}{\Phi}_j}{\partial y^{2}} 
= -\frac{g^2}{2\text N\text T^2} \frac{\partial{V_{eff}}}{\partial{{\Phi}_j}} ;
\quad j = 1, 2
\end{equation}
with $\Phi=\Phi_1+i\Phi_2$
The evolution of the field was numerically implemented by
a stabilized leapfrog algorithm of second order accuracy
both in space and in time with the second order derivatives
of $\Phi$ approximated by a diamond-shaped grid. Here, we have used a square 
lattice of $2000 \times 2000$ points and the physical size of the lattice is 
20 fm. Hence, the lattice 
spacing is $\Delta x=0.01$ fm. We
take $\Delta t = \Delta x/\sqrt 2$ to satisfy the
Courant stability criteria. The stability and accuracy of the
simulation is checked using the conservation of energy during
simulation. The total energy fluctuations remains few percent
without any net increase or decrease of total energy in the
absence of the dissipative term $\dot{\Phi}$ in the equation of motion. This is 
the only dissipative term used here due to Bjorken's longitudinal expansion. Here
the temperature varies as $ \tau^ {-1/3}$ due to Bjorken's longitudinal expansion.   

It has been already shown that for very small explicit symmetry breaking term with
$b_1$=0.005 (for a first order phase transition), Z(3) domains are formed 
via bubble nucleation, then expand and Z(3) walls and strings are 
produced \cite{gupta2}. Z(3) domains also have been studied in a quench scenario
with this symmetry breaking term, where it has been seen that true vacuum domain 
dominates over other two Z(3) metastable domains\cite{ranj}. However, in this work 
as mentioned in the previous section, we consider the case with $b_1=0.1$ which is
suitable for a crossover transition between the confined and deconfined phase at zero
chemical potential. With this symmetry breaking term, the plot of the potential 
along true and metastable vacua is given in Fig.7. As we have mentioned 
earlier, the potential along metastable vacua is higher compared to 
true vacuum near the origin $\Phi=0$. So, if we 
choose the initial patch for the Polyakov loop field near $\Phi=0$ to be very small i.e
$\epsilon = 0.01\times vev$, then the field always rolls down to the true vacuum. In this case, 
we don't get any Z(3) domain structure in the simulation. 

The patch size for
the initial configuration of the Polyakov loop depends upon the flatness of the
potential near $\Phi$ =0 (Polyakov correlation length) at temperature less than critical
temperature. The initial Polyakov loop field can take
large value for a flat potential near $\Phi=0$ compared to a narrow potential. We choose
a larger initial patch i.e $\epsilon = 0.1 \times vev$ for this potential, the field already
sits on the top of the barrier along the metastable vacua near $\Phi=0$. Then the field rolls
down to all vacua, since the potential is more tilted towards the true vacuum,
the larger fraction of region is occupied by domains of true vacuum inside as
shown in Fig.10. Here, we have shown the
values of the phase of $\Phi$  around the three Z(3) vacua in terms of different colors (shades)
to focus on the evolution of Z(3) domain structure. Thus all
the values of the phase $\theta$ of $\Phi$ are separated in three ranges,
between -2$\pi/6$ to $2\pi/6$ ($\theta$ = 0 vacuum), between $2\pi/6$ to $\pi$
($\theta$ = $2\pi/3$ vacuum), and between $\pi$ to − 2$\pi/6$ ($\theta$ = $4\pi/3$
vacuum). As the field evolves, the angular variation
of $\Phi$ becomes less random over small length scales, leading
to a sort of Z(3) domain structure. Z(3) domains become
more well defined, and grow in size by coarsening as shown
in sequence of figures in Fig.10.
The boundaries of different Z(3) domains represent Z(3)
walls, and the junction of three different Z(3) domains give
rise to the QGP strings.

Here, we would like to mention that at temperature 400 MeV as shown in 
Fig.7, the potential along the metastable vacuum 
has a small barrier between confined vacuum and metastable vacuum. So, there is a 
possibility of metastable domain formation via bubble nucleation. Since the 
barrier height is much smaller than energy density difference between the confined 
and metastable vacua, these will be thick wall bubbles.
But, there is no barrier between confined vacuum and true vacuum at this 
temperature and true vacuum domains are formed due to the roll down of the Polyakov 
loop field. Hence, this raises a new possibility of Z(3) domain formation where 
the transition dynamics are different along different vacua. We would 
like to study this in detail in the future. 

\begin{figure*}[!hpt]
\begin{center}
\leavevmode
\epsfysize=6truecm \vbox{\epsfbox{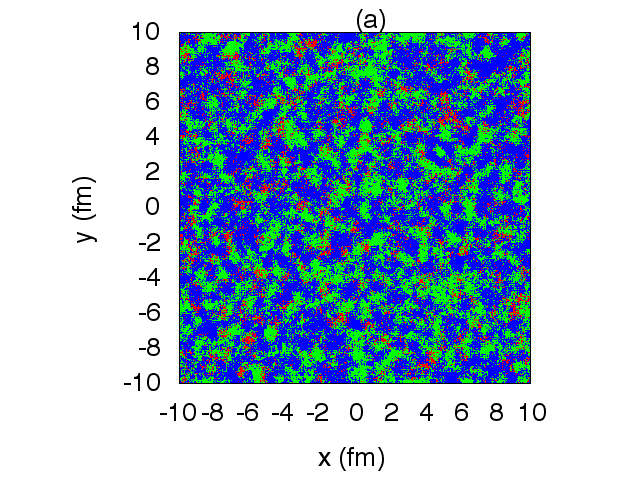}}
\epsfysize=6truecm \vbox{\epsfbox{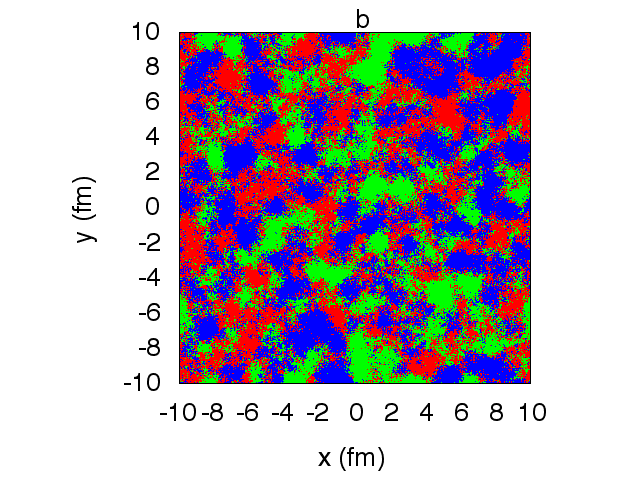}}
\epsfysize=6truecm \vbox{\epsfbox{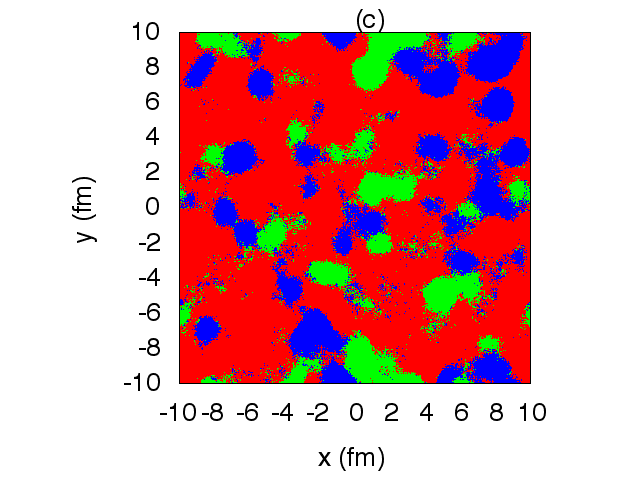}}
\epsfysize=6truecm \vbox{\epsfbox{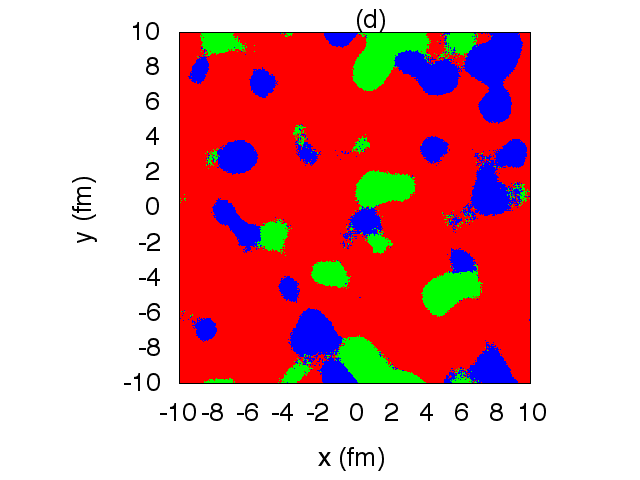}}
\end{center}

\caption{}{Field configurations at different times with explicit symmetry breaking effect. 
The shading (color) representing the dominant region in (d) corresponds to the true 
vacuum with $\theta$ = 0. (a)–(d) show the growth of domains for $\tau$ = 1.2, 1.6, 2.2,
and 2.4 fm (with corresponding values of temperature T = 368, 335, 311 and 298 MeV respectively.}
\label{Fig.10}
\end{figure*}

Next, we discuss the situation for PQM model. The plots of the potential
for this model along true and metastable vacua at T = 400 MeV has been shown in Fig.7. It is
clear from the figure that the energy density difference between true and metastable
vacua of this model is very large compared to the previous case with $b_1$=0.1. 
Since it is a large explicit symmetry breaking case for $V_{eff}$, with 
the initial field configuration mentioned above for $b_1$=0.1 (i.e $\epsilon=0.1\times vev$
, the field $\Phi$ always rolls down completely along $\theta$ = 0 vacuum. There are 
no metastable domain formation for this 
case. When the field always rolls down to true vacuum, it has been observed that
large field oscillations lead to large fluctuations in the
evolution of flow anisotropies in quench case compared to equilibrium 
transition case \cite{ranj}. We expect similar results for this case as in \cite{ranj}.
Here, the order parameter $\sigma$ is a dynamic field which evolves as the Polyakov 
loop field in the quench scenario. The initial field configuration for $\sigma$ is 
chosen appropriately ( vev value $\sim$ 93.0 MeV at zero temperature and sigma value 
fluctuating around this value from one lattice site to another). Equation of motion for 
the two order parameters are now coupled through $V_{eff}$. While the dynamical 
equation for the Polyakov loop order parameter is given
by Eq.(\ref{eomphi}), the  equation of 
motion for $\sigma$ field is given by

\begin{equation}
\label{eomsig}
  \frac{\partial^{2}{\sigma}}{\partial\tau^{2}} + \frac{1}{\tau} 
\frac{\partial{\sigma}}{\partial \tau}  -\frac{\partial ^{2}{\sigma}}
{\partial x^{2}}   
-\frac{\partial^{2}{\sigma}}{\partial y^{2}} 
= -\frac{\partial{V_{eff}}}{\partial{\sigma}} 
\end{equation}

Here we would like to study true vacuum and sigma domain growth in a non central
collision. It is a limitation of current study to use hydrodynamic simulation. 
In contrast to the previous case, here we use a more realistic temperature profile of 
Woods-Saxon shape with the size in the X and Y
directions being different representing elliptical shape for a
non central collision. The Woods-Saxon temperature profile is given by the following

\begin{equation}
\text T(r)=\text T_{center}/[1+exp[(r-R)/a]]
\end{equation}

where $\text T_{center}$ is the temperature at the center of the QGP region. R represents 
the radius of the elliptical QGP region for the non-central collision and a=0.56 fm is 
the thickness of transition layer to the vacuum. Here r represents the distance from the
center of the lattice at which temperature is measured using the above equation.
 At $\tau =1 fm $ (thermalization time scale used here), the temperature at the 
center of QGP region is $\text T_{center}= 400 MeV$. 
This allows us to have a well defined
size for the central QGP region, with temperature smoothly
decreasing at the boundary of this region \cite{ranj}. The transverse
size R of this system is taken to increase with uniform acceleration of 0.015 
c per fm, starting from an initial value of R \cite{kolb}.
The initial transverse expansion velocity is taken to be zero.
This expanding background of temperature profile is supposed
to represent the hydro-dynamically expanding quark gluon plasma.
The central temperature of the Woods-Saxon profile is taken to
decrease by assuming that the total entropy (integrated in
the transverse plane) decreases linearly as appropriate for
Bjorken dynamics of longitudinal expansion. The physical lattice size is
$30 fm \times 30 fm$. The Woods-Saxon temperature profile
(representing QGP region) is taken to have a diameter of about
16 fm as appropriate for Au-Au collision for RHICEs. The
large physical size of the lattice allows for the evolution of the
QGP region to be free from boundary effects. Here we have taken eccentricity
of QGP region to be 0.8. We have presented the Polyakov loop 
and sigma domain growth at time $\tau$ = 3 fm in Fig.11. Fig.11a represents
true vacuum domain and Fig 11.b presents sigma domain growth at $\tau$ = 3 fm 
respectively. Here the ring like structure at the boundary of non central QGP
region represents the large fluctuations in the field value due to huge 
temperature gradient corresponding to the tail part of Woods-saxon 
temperature profile. However, we are only interested in the domain growth
inside the QGP region ( neglecting the boundary effects). The background is 
in the confined phase (outside of the 
elliptical QGP region). So, the Polyakov loop field and sigma field value 
outside QGP region is accordingly in the confined phase.   

\begin{figure*}[!hpt]
\begin{center}
\leavevmode
\epsfysize=6truecm \vbox{\epsfbox{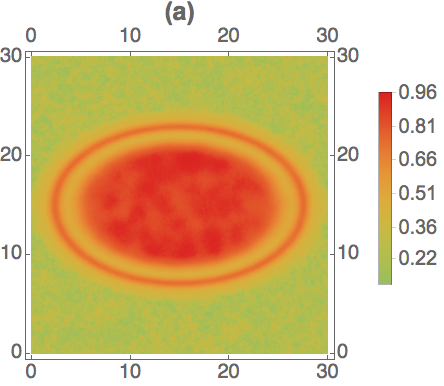}}
\epsfysize=6truecm \vbox{\epsfbox{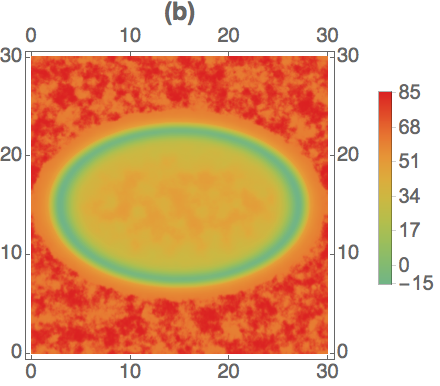}}
\end{center}
\caption{}{(a) Polyakov loop (b) sigma field (in units of MeV) domain growth at $\tau$ = 3 fm}
\label{Fig.11}
\end{figure*}

We also study the variation of average Polyakov loop and sigma field (normalized with
respect to vacuum) over the 
whole lattice with $\tau$ as shown in Fig.12. This is very important to study
the relaxation behavior of the Polyakov loop and sigma field in a quenched 
scenario. The average Polyakov loop and sigma field over the lattice is defined 
as
\begin{equation}
\left<|\Phi|\right>=\frac{1}{N^2}\sum_{l,m}|\Phi|_{l,m} \quad and \quad 
\left<\sigma\right>= \frac{1}{N^2}\sum_{l,m}\sigma_{l,m}. 
\end{equation}

Here N represents the number of 
lattice points in one direction and (l,m) represents the coordinates 
of the lattice. We can observe 
from Fig.12 that the average Polyakov loop initially
increases up to $\tau= 2$ fm approaching the equilibrium value, however, 
at the same time, temperature decreases with $\tau$ due to longitudinal expansion 
and the equilibrium value of the Polyakov decreases as temperature decreases. Hence,
there is a turn over after $\tau =2 fm$ where the Polyakov loop decreases. This is also 
the case for normalized sigma field value, where it initially decreases approaching 
the equilibrium value and increases slightly after $\tau=2 fm$. We would like to point 
out that the central temperature is $\text T_{center}= 313 MeV$ at $\tau=2 fm$  and there 
is a large fraction of the lattice which is at low temperature due to Woods-Saxon 
temperature profile. We would also like to point out that as the temperature decreases
further at $\tau = 3 fm$, the shape of the effective potential is very flat near the minimum. 
So, the equilibrium value of the Polyakov loop oscillates slightly near the vev which is shown as
increase in the Polyakov loop around $\tau= 3 fm $ 
   
\begin{figure*}[!hpt]
\begin{center}
\leavevmode
\epsfysize=6truecm \vbox{\epsfbox{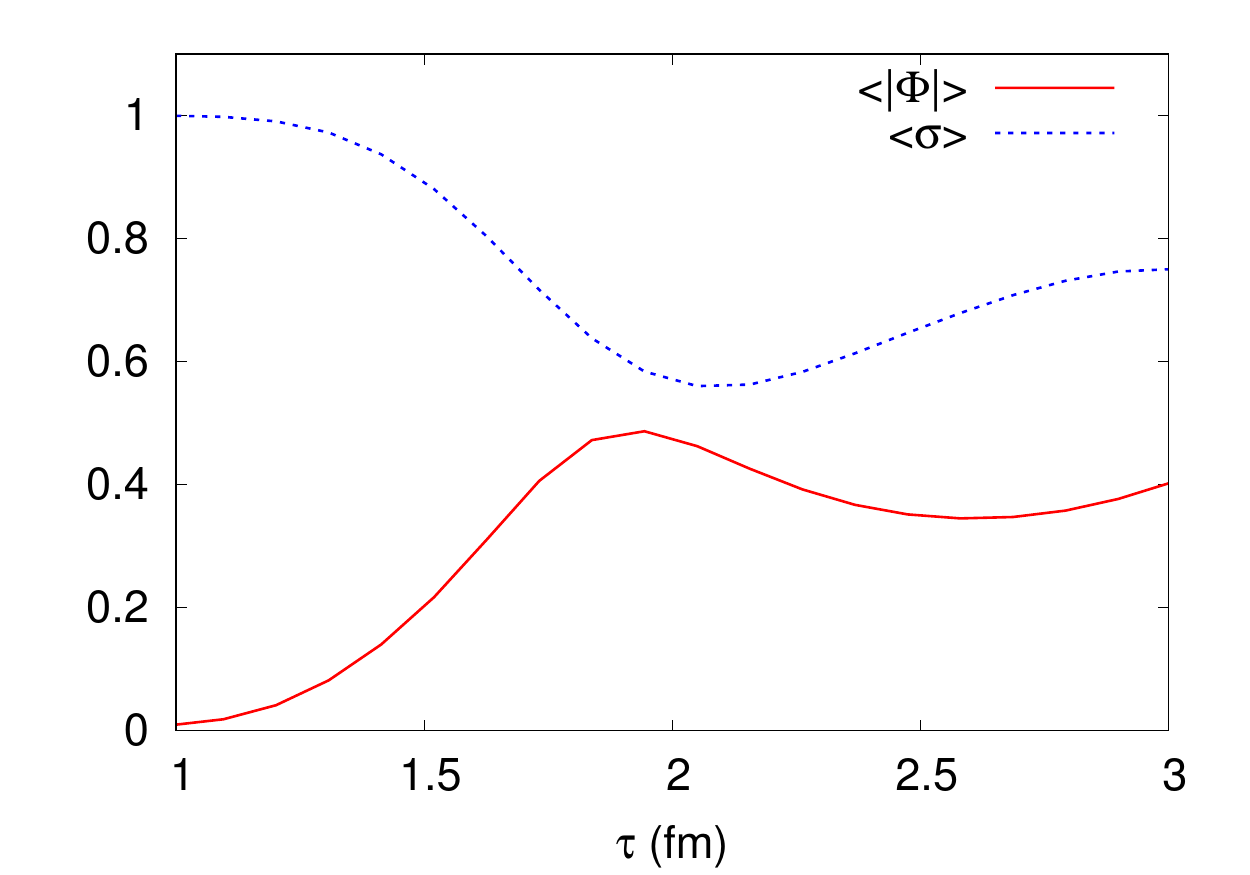}}
\end{center}
\caption{}{Plot of average Polyakov loop and normalized sigma field value with $\tau$.}
\label{Fig.12}
\end{figure*}

As we have described above for $b_1$=0.1, there is a possibility of Z(3) metastable 
domain formation via bubble nucleation due to the barrier between false 
vacuum and Z(3) metastable vacuum for this case in PQM model. However, here the barrier
height is of the same order as the energy density difference between confined vacuum
 and metastable vacua at a temperature 400 MeV as shown in Fig.7. But, there is no 
barrier between false vacuum and true vacuum at this temperature. So, there is a possibility 
of Z(3) domain formation in this model via mixed transition dynamics i.e roll down 
of the Polyakov loop field along true vacuum and bubble nucleation along 
metastable vacua. One needs to calculate the nucleation probability for this 
bubble formation and to study the phase transition dynamics 
with a suitable bubble profile as in \cite{gupta,gupta2}. We would 
like to study this case in future.

\section{CONCLUSIONS}

We have presented about the possibility of existence of Z(3) metastable states in PQM model.
The metastable states exist for the Polyakov loop potential with large barrier 
between different Z(3) vacua. The metastable vacua are not present near the 
critical temperature, but they appear around temperature 310 MeV at zero chemical
potential due to strong explicit symmetry breaking effect of quarks. There is also 
a shift in the phase of metastable vacua and they appear $115^{\circ}$ and 
$245^{\circ}$) respectively. This explicit symmetry breaking effect in PQM model 
is strong compared to small explicit breaking effect of quarks discussed in different
Polyakov loop models \cite{psrsk2}. We have observed that this metastability temperature 
remains almost constant up to finite chemical 170 MeV, and it increases after this chemical 
potential. This might be due to the difference in phase transition dynamics after 
the critical chemical potential 170 MeV. The phase transition between confined and
deconfined phase is a crossover up to the critical chemical potential and it is a first
order phase transition beyond this chemical potential. In other words, one can say that
the metastability temperature as a function of chemical potential is sensitive to
the phase transition dynamics.

It has been suggested that Z(3) domains give a microscopic explanation for large color 
opacity ( jet quenching) and near perfect fluidity ( small value of $\eta/s$) nature 
of QGP \cite{bass,monnai}. Therefore, it is very important to study these domains near critical
temperature in effective models like PQM or Polyakov loop extended Nambu-Jona-Lasinio (PNJL) model. We expect the  results will be similar in 
PNJL model. Here we have observed in a quench scenario that due to strong explicit 
symmetry breaking of quarks in PQM model, there are no metastable domain formation. 
The whole region is occupied by only true vacuum domain. Hence, it is very unlikely
to explain jet quenching or perfect fluidity nature of QGP due to Z(3) domains in 
PQM model. However, this nature of QGP can be described well in models with small 
explicit symmetry breaking effect of quarks.        

\acknowledgments

We are very grateful to Ajit M Srivastava for very useful discussions and suggestions.
We also thank Arpan Das and Ananta P Mishra for useful discussions. We would
like to thank Ananta P Mishra for providing the code of energy minimization to get the
interface profile.    

%%%%%%%%%%%%%%%%%%% 

\end{document}